# Tuning thermoelectric properties of graphene/boron nitride heterostructures

Laith A. Algharagholy[†a,b,c,d], Qusiy Al-Galiby[†b,c,f], Haider A. Marhoon[†e], Hatef Sadeghi[†b,c], Hayder M. Abduljalil[†e], Colin J. Lambert[†b,c*]

[a]College of Computer Science and Mathematics, Al-Qadisiyah University, Diwaniyah, Iraq.
[b]Department of Physics, Lancaster University, Lancaster LA1 4YB, United Kingdom.
[c]Quantum Technology Centre, Lancaster University, LA1 4YB Lancaster, UK.
[d]College of Basic Education, Sumer University, Al-Refayee, Thi-Qar, Iraq.
[e]Department of Physics, Babylon University, Babylon, Iraq.
[f]Department of Physics, Al Qadisiyah University, Al Qadisiyah, Iraq.
†Authors contributed.
*Corresponding author: c.lambert@lancaster.ac.uk

## Abstract

Using density functional theory combined with a Green's function scattering approach, we examine the thermoelectric properties of hetero-nanoribbons formed from alternating lengths of graphene and boron nitride. In such structures, the boron nitride acts as a tunnel barrier, which weakly couples states in the graphene, to form mini-bands. In un-doped nanoribbons, the mini bands are symmetrically positioned relative to the Fermi energy and do not enhance thermoelectric performance significantly. In contrast, when the ribbons are doped by electron donating or electron accepting adsorbates, the thermopower $S$ and electronic figure of merit are enhanced and either positive or negative thermopowers can be obtained. In the most favourable case, doping with the electron donor tetrathiafulvalene (TTF) increases the room-temperature thermopower to -284 μv/K and doping by the electron acceptor tetracyanoethylene (TCNE) increases $S$ to 210 μv/K. After including both electron and phonon contributions to the thermal conductance, figures of merit $ZT$ up to of order 0.9 are obtained.

## 1. Introduction

The ability to manage waste heat is a major challenge, which currently limits the performance of information technologies. To meet this challenge, there is a need to develop novel materials and device concepts, innovative device architectures, and smart integration schemes, coupled with new strategies for managing and scavenging on-chip waste heat. The development of new high-efficiency and low-cost thermoelectric materials and devices is a major target of current research. Thermoelectric materials, which allow highly-efficient heat-to-electrical-energy conversion from otherwise wasted low-level heat sources, would have enormous impact on global energy consumption.

Nanoscale systems and especially nanoscale structures are very promising in this respect, due to the fact that transport takes place through discrete energy levels. The ability to measure thermopower in nanoscale junctions opens the way to developing fundamentally-new strategies for enhancing the conversion of heat into electric energy. The thermopower (or Seebeck coefficient) $S$ of a material or nanoscale device is defined as $S = -ΔV/ΔT$, where $ΔV$ is the voltage difference generated between the two ends of the junction when a temperature difference $ΔT$ is established between them. In addition to the goal of maximising $S$, there is a world-wide race to develop materials with a high thermoelectric efficiency $η$. This is often expressed in terms of a dimensionless figure of merit[1-3] $ZT = (S^2G/κ)T$, where $S$ is the Seebeck coefficient, $G$ is the electrical conductance, $T$ the temperature and $κ$ the thermal conductance given by $κ = κ_e + κ_p$, where $κ_e$ ($κ_p$) is the electronic (phononic) contribution to $κ$. In terms of $ZT$, the maximum efficiency of a thermoelectric generator is $η_{max} = η_c (α-1)/(α+1)$ where $η_c$ is the Carnot efficiency and $α = (ZT+1)^{1/2}$, whereas the efficiency at maximum power is $η_p = η_{CA} (α^2-1)/(α^2+1)$, where $η_{CA}$ is the Curzon-Ahlborn upper bound. In both cases, the efficiency is a maximum when $ZT$ tends to infinity.

A key strategy for improving the thermoelectric properties of inorganic materials has been to take advantage of nanostructuring[4-10], which leads to quantum confinement of electrons, suppression of parasitic phonons[11, 12] and enhanced thermoelectric performance[13, 14]. For example, nanostructured materials such as PbSeTe/PbTe-based quantum dot superlattices with *ZT* of ca. 2 were realized over a decade ago[5]. However lack of further improvement since that time suggests that new strategies are needed. More importantly, none of the above materials use sustainable elements. The raw $Bi_2Te_3$ material is already globally limited (<300 ton yr$^{-1}$) and long term resources of tellurium are close to exhaustion. Furthermore tellurium is toxic and energetically expensive to process.

Recently the thermoelectric properties of graphene have been investigated for their potential to convert waste heat into electricity[1-3, 15-21]. The challenge of maximising *ZT* involves maximising the power factor $S^2GT$ and minimising the thermal conductance $\kappa$. The thermal conductance $\kappa = \kappa_e + \kappa_p$ is the sum of the electronic ($\kappa_e$) and phononic ($\kappa_p$) thermal conductances. Since the task of minimising the latter involves phonon engineering, whereas enhancement of $S^2GT$ and minimisation of $\kappa_e$ involves the tuning of electronic properties, these two tasks are often considered separately.

As a first step one should identify materials in which the power factor and electronic figure of merit $ZT_e=S^2GT/\kappa_e$ are maximized and as a second step, if these properties are favourable, develop strategies for minimizing $\kappa_p$. In what follows, we tackle the former challenge by developing new strategies for optimizing the thermoelectric performance of graphene-boron nitride nanoribbons. Intuitively our strategy can be understood by examining the transmission coefficient *T(E)* of electrons of energy *E* passing through a nanostructure from a hot to a cold reservoir. If the Fermi energy of the reservoirs is $E_F$, then it is convenient to introduce the variable *x = (E - $E_F$)*. If the distribution of *x* is *ρ(x)*, then the mean of *x* is <x> = ∫dx ρ(x)x and the variance is $\sigma^2$ = <$x^2$> - <x>$^2$ = ∫dx ρ(x)(x- <x>)$^2$. In terms of these quantities, the Seedbeck coefficient (or thermopower) *S* is given by

$$S = \frac{1}{eT} <x>, \quad (1)$$

where *T* is the temperature and *e* is the electronic charge. Similarly the electronic contribution to the figure of merit $ZT_e$ is

$$ZT_e = \frac{<x>^2}{\sigma^2}. \quad (2)$$

(see below for a derivation of these equations.) Clearly $ZT_e = \infty$, when $\sigma = 0$, which means that $\rho(x)$ should be proportional to a delta function[10] of the form $\rho(x) = A\delta(x - x_0)$, in which case $S = \frac{1}{eT}Ax_0$. This strategy is relevant for structures, whose electronic density of states contains narrow resonances, such as single-molecule electrical junctions[22, 23]. However in what follows, we shall find that BN-graphene nanoribbons do not exhibit narrow resonances and therefore an alternative strategy is needed.

## 2. Thermoelectric properties of BN-graphene hetrostructures

Graphene and boron nitride are attractive, because of their unique mechanical and electrical properties and their closely-matched lattice parameters[24-26]. To develop a strategy for optimizing their thermoelectric properties, we first relate ρ(x) to physical quantities by introducing the non-normalized probability distribution *P(E)* defined by [27, 28]

$$P(E) = -T(E)\frac{\partial f(E)}{\partial E}, \quad (3)$$

where *T(E)* is the electron transmission coefficient and *f(E)* is the Fermi distribution function. The moments of *P(E)* are

$$L_n = \int_{-\infty}^{\infty} dE\, P(E)(E - E_F)^n, \quad (4)$$

and the electrical conductance, *G* is given by the Landauer formula

$$G = \frac{2e^2}{h}L_0, \quad (5)$$

where *h* is Planck's constant. In terms of these moments, the thermopower is $S = \frac{1}{eT}\frac{L_1}{L_0}$, the electronic contribution to the thermal conductance is

$$k_e = \frac{2}{h}\frac{1}{T}(L_2 - \frac{L_1^2}{L_0}) \qquad (6)$$

and

$$ZT_e = \left(\frac{S^2 G}{k_e}\right)T = \left[\frac{L_1^2}{L_0^2}\right]/\left[\frac{L_2}{L_0} - \frac{L_1^2}{L_0^2}\right]. \qquad (7)$$

To derive equations (1) and (2), we define the normalised distribution[22] $\rho(x)$ by $\rho(x)=P(x)/L_0$. In terms of the moments $L_n$, the mean and standard deviation of $\rho(x)$ are given by $<x> = \frac{L_1}{L_0}$, $\sigma^2 = \frac{L_2}{L_0} - \frac{L_1^2}{L_0^2}$. Substituting these into equations (6) and (7) yields equations (1) and (2) and furthermore, the electronic contribution to the thermal conductance ($k_e$) becomes

$$k_e = \frac{2}{h}\frac{1}{T}L_0\,\sigma^2, \qquad (8)$$

Clearly $<x>$ and $\sigma^2$ capture essential features regarding the shape of $\rho(x)$ and $P(x)$. These shape-dependent parameters are independent of the normalisation constant $L_0$ of $P(x)$. Equations (1) and (2) reveal that $S$ and $ZT_e$ depend only on these shape parameters and are independent of $L_0$. Only the electrical and electronic thermal conductances depend on $L_0$. This feature which can be traced to the fact that $G$ and $k_e$ describe the magnitudes of currents and therefore depend on the magnitude of $P(x)$, whereas $S$ and $ZT_e$ involve only ratios.

Since $[-\frac{\partial f(E)}{\partial E}]$ is an even function of $x$ of width $k_B T$, our strategy will involve engineering BN-graphene nanoribbons to create an asymmetry in $T(E)$, located close $E_F$. In what follows, we shall find that it is possible to create sharp steps in the electron transmission coefficient $T(E)$ of BN-graphene nanoribbons and therefore to understand this strategy, it is of interest to examine the thermopower and $ZT_e$ of a system with a model transmission coefficient of the form: $T(E) = A$ for $E<E_0$ and $T(E)=0$ for $E>E_0$, where $A$ is an arbitrary constant defining the height of the step and $E_0$ defines the position of the step[21]. In this case, since the only energy scale is $k_B T$, where $k_B$ is Boltzmann's constant and $T$ is the temperature, it is convenient to introduce the dimensionless parameter $y = (E - E_F)/k_B T$, so the Fermi function takes the form $f(E)=(\exp y + 1)^{-1}$.

In this case, equation (4) yields

$$L_n = A(k_B T)^n I_n(y_0) \qquad (9)$$

where $y_0 = (E_0 - E_F)/k_B T$,

$$I_n(y_0) = \int_{-\infty}^{y_0} dy\, P(y) y^n \qquad (10)$$

and $P(y) = -df/dy = e^y/(e^y + 1)^2$. Clearly all moments depend only on the size of the step (ie the dimensionless parameter $A$) and the dimensionless parameter $y_0$, which defines the location of the step relative to the Fermi energy of the electrodes, in units of $k_B T$. In terms of $I_n$, $<x> = k_B T \frac{I_1}{I_0}$, $\sigma^2 = (k_B T)^2 [\frac{I_2}{I_0} - \frac{I_1^2}{I_0^2}]$. In terms of the dimensionless Fermi-function integrals $I_n(y_0)$, the thermoelectric parameters become

$$G = \frac{2e^2}{h} A I_0, \qquad (11)$$
$$S = \frac{k_B}{e}\frac{I_1}{I_0}, \qquad (12)$$
$$k_e = \frac{2A(k_B)^2 T}{h}(I_2 - \frac{I_1^2}{I_0}), \qquad (13)$$
$$ZT_e = \left[\frac{I_1^2}{I_0^2}\right]/\left[\frac{I_2}{I_0} - \frac{I_1^2}{I_0^2}\right] \qquad (14).$$

These equations show that the natural unit of $G$ is $\frac{2e^2}{h} = 77\,\mu S$, of $S$ is $\frac{k_B}{e} = 86\,\mu V/K$ and of $k_e$ is $\frac{2(k_B)^2 T}{h} = 173\,pW/K$ at room temperature (ie 300K). Plots of the resulting thermoelectric parameters are shown in figure 1. Clearly $G$ and $k_e$ are both proportional to step size $A$, whereas $S$ and $ZT_e$ are independent of $A$.

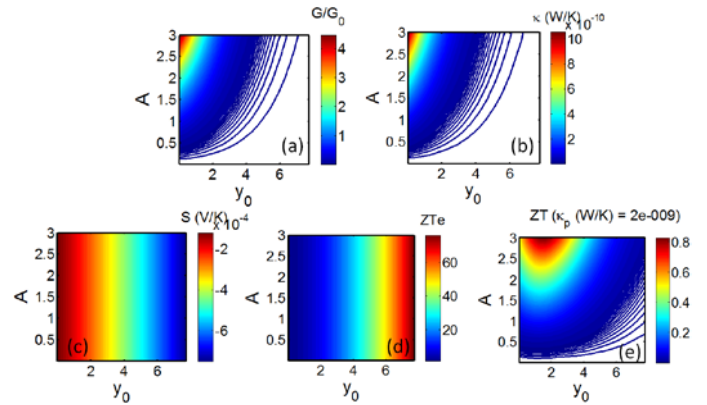

**Figure 1.** Plots of thermoelectric properties of a system with a model step-like transmission coefficient[21], versus the dimensionless parameter $y_0$, which defines the location of the step. (a) electrical conductance, (b) electronic contribution to the thermal conductance $k_e$, (c) Seebeck coefficient $S$, (d) electronic contribution to the figure of merit $ZT_e$ and (e) full figure of merit at ($k_p$= 2 nW/K) as a function of the position of step function $y_0$ and the amplitude of T(E).

In what follows, we shall find that it is possible to create sharp steps in $T(E)$ of BN-graphene nanoribbons, but these are not necessarily located in optimal positions relative to $E_F$. Therefore to optimize $S$ and $ZT_e$ we shall tune the asymmetry by doping. To avoid introducing unnecessary scattering, rather than introducing substituents, we shall consider doping the ribbons using adsorbates[10] of the electron acceptor tetracyanoethylene (TCNE)[29, 30] and the electron donor tetrathiafulvalene (TTF)[30, 31].

To benchmark our study, we first examined electron transport in the five zigzag ribbons shown in figure S8 of the SI, which are infinitely periodic in the longitudinal $z$ direction and contain narrow transverse strips of BN, separated by strips of graphene of length $l$=1, 2, 3, 4 and 5 hexagons. The $l=2$ structure is shown in figure 2.

We shall find that the $T(E)$ of these structures is symmetric about $E_F$, and therefore their thermopowers are low. To break this symmetry, we shall dope the materials using the organic molecules TCNE and TTF, to yield the structures shown in figure S9 and S10 of the SI, examples of which (for $l=3$) are shown in figures 3 and 4. In all cases, we use supercells containing five unit cells, to allow for incommensurabilities following relaxation.

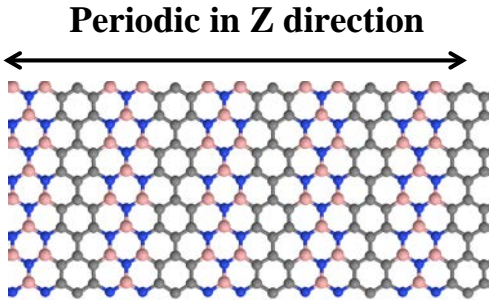

**Figure 2:** Shows the relaxed supercell of a $l=1$ graphene/boron nitride hetero-ribbon, with 240 carbon, boron and nitrogen atoms.

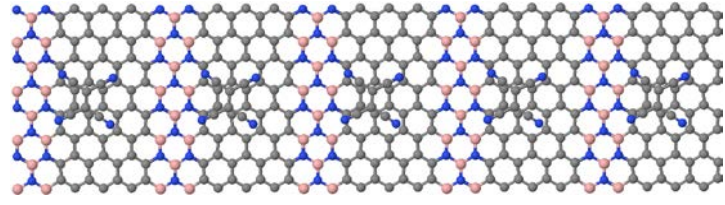

**Figure 3.** Examples of a TCNE-doped structure. The figure shows the DFT calculated optimum geometry for the $l=3$ structure the adsorbed electron-acceptor TCNE.

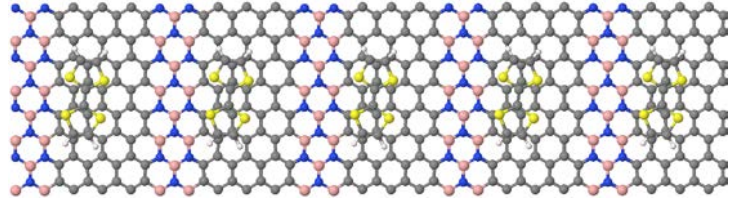

**Figure 4.** Examples of a TTF-doped structure. The figure shows the DFT calculated optimum geometry for the $l=3$ structure the adsorbed electron-donor TTTF.

## 3. Computational Method

To obtain the optimised geometry (and the associated mean-field Hamiltonian) of the above structures, we employed the *SIESTA*[32] implementation of Density Functional Theory (DFT). The local density approximation (LDA) with norm-conserving pseudopotentials, double-zeta polarized (DZP) basis sets of pseudoatomic orbitals, a real-space grid defined with a plane-wave cut-off energy of 250 Ry, and the Ceperley-Alder exchange (CA) correlation functional with the atomic forces relaxed to 0.02 eV/˚A have been used in this work. The underlying mean-field Hamiltonian obtained from *SIESTA* was combined with the *GOLLUM*[33], implementation of the non-equilibrium Green's function (NEGF) method to find the transmission coefficient $T(E)$ for electrons with energy $E$ passing from one electrode to another.

## 4. Results and discussion

Figure 5 shows the calculated band structures (right-hand plots) and the number of open scattering channels (left-hand plots) of the five structures shown in figure S8. The latter are plotted as a function of electron energy $E$, relative to the DFT-predicted value of the Fermi energy $E_F^{DFT}$. To understand these trends, the BN should be considered as a tunnel barrier, which weakly couples states in the graphene islands to form mini-bands. As the size of the graphene regions is increased, the spacing between the quantised states decreases and the number of mini-bands entering the BN energy gap increases. As shown by the left-hand plots of figure 5, each mini-band contributes a scattering channel and since the transmission coefficient $T(E)$ of such periodic structures is equal to the number of open channels at energy $E$, this leads to step-like features in $T(E)$ at the mini-band energies.

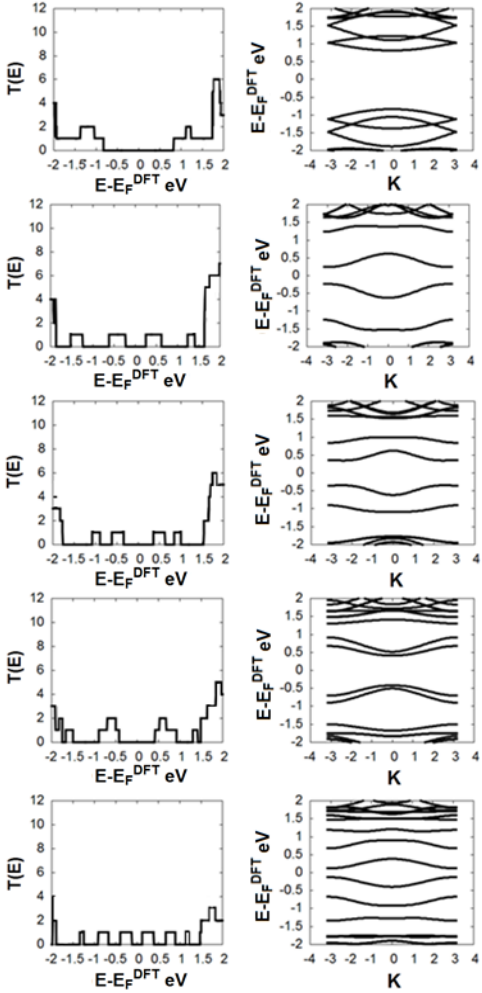

**Figure 5:** Band structures (right) and $T(E)$ (left) for the structures of figure 2. $E_F^{DFT}$ is the DFT-predicted value of the Fermi energy.

Although these features are desirable, their symmetric location relative to the Fermi energy means that they are unlikely to improve thermoelectric performance. Therefore it is of interest to examine how properties would change if the true Fermi energy ($E_F$) were allowed to vary from the DFT-predicted value ($E_F^{DFT}$). In practice the Fermi energy can be varied experimentally by electrostatically or electrochemically gating the wire or by doping. To this end, we now consider the effect of doping with TTF and TCNE and perform fully self-consistent calculations which take into account changes in the total charge of the system. A first step in the calculations is to relax the structures in the presence of adsorbates, to yield structures such as those shown in figure 6. In each case after geometry relaxation, the binding energy and charge transferred from the two molecules were computed as shown in table 1. These calculations show that TCNE gains electrons from the surface of graphene-boron nitride, whereas the TTF donates electrons to the surface.

**Table 1:** This shows DFT calculations of the charge transferred and binding energies between the surface of graphene-boron nitride and the two adsorbates TCNE and TTF.

|  | UNITa | | UNITc | | UNITe | |
| --- | --- | --- | --- | --- | --- | --- |
|  | TCNE | TTF | TCNE | TTF | TCNE | TTF |
| $\Delta E$ (eV) | -0.976 | -0.554 | -1.157 | -0.957 | -0.814 | -0.617 |
| $\Delta N$ (e) | 0.054 | -0.172 | 0.251 | -0.47 | 0.37 | -0.497 |

As an example, Figure 6 shows side views of the optimised structures of ribbon (a) after relaxation. Fig. 5II and 5III show the optimized graphene-boron nitride structure when doped by TCNE and TTF. In these structures, TCNE π-stacks a distance *0.28 nm* above the surface, while the TTF π-stacks at a distance *0.3 nm*. As an example, figure 6 shows the structure-e (I) without doping, (II) doped by electron acceptor-TCNE, and (III) doped by electron donor-TTF.

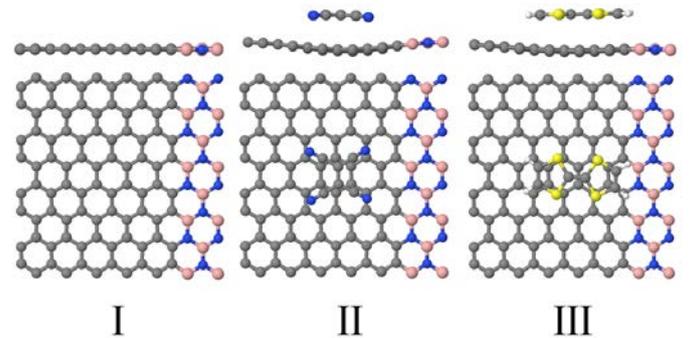

**Figure 6:** Side and top views of optimised structure-a (1BN-5G) for molecular complexs: (I) without doping, (II) doped by electron acceptor-TCNE, and (III) doped by electron donor-TTF.

After obtaining all relaxed structures, the transmission coefficient $T(E)$ (= number of open channels) was obtained from their mini-band structures and thermoelectric coefficients calculated using equations (3) to (6). For structure (e) of figures S8-S10 of the SI, figure 7 shows results for transmission function $T(E)$ (a) without doping, (b) doped by the electron acceptor TCNE and (c) doped by the electron donor TTF, along with results for the thermopower $S$ and the full figure of merit $ZT$. This demonstrates that a Fermi energy shift can be realised by introducing donor or acceptors onto the surface of the graphene-boron nitride.

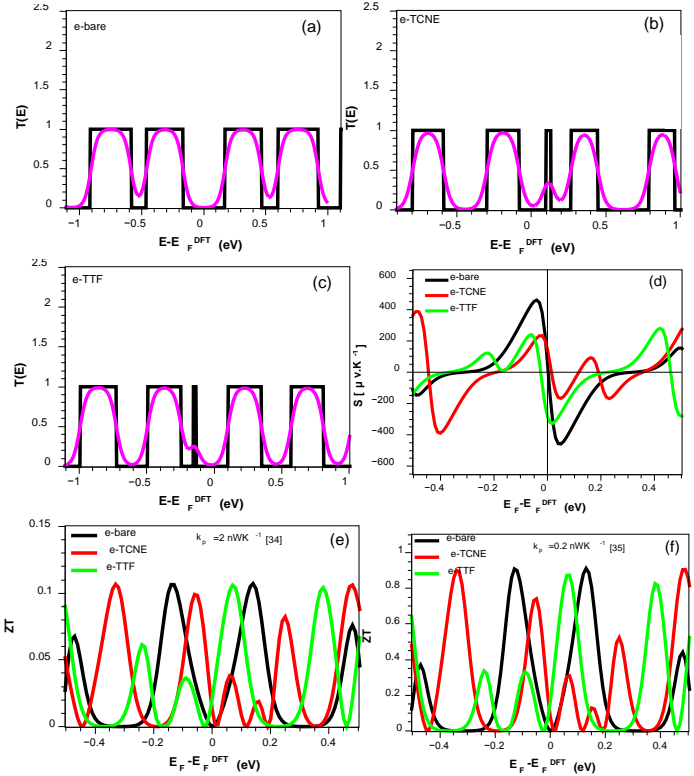

**Figure 7:** For the graphene-boron nitride structure (e) of figure S9, the black solid line in each panel shows DFT calculations of the transmission coefficient $T(E)$ for electrons of energy $E$ passing from one electrode to the other (a) without doping, (b) doped by TCNE and (c) doped by TTF. The magenta lines show the room temperature electrical conductance. (color online) Figures 7d, 7e and 7f show the room-temperature thermopower $S$, the figure of merit $ZT$, obtained by assuming a phonon thermal conductance of 2nW/K [34] and 0.2nW/K [35]. Since the Fermi energy $E_F^{DFT}$ predicted by density functional theory is not necessarily accurate, Figs (d-f) show results for $S$ and $ZT$ obtained from equ. (4) using different values of $E_F$.

Figures 7 and 8 show corresponding results for the structures (c) and (a) of figure S8. Figures 7a, 8a and 9a show that as the length of the graphene regions increases, an increasing number of mini-bands are formed from quantized levels within the graphene, leading to an increasing number of rectangular steps in $T(E)$. However none of the un-doped structures possess sharp asymmetrically-located features in $T(E)$. On the other hand, figures 7b, 8b and 9b (7c, 8c and 9c) show that the presence of TCNE (TTF) adsorbates creates a narrow feature placed asymmetrically near $E_F^{DFT}$. This delta-function-like feature[10] is associated with the interaction between localized states on the analytes and extended states on the surface of graphene-boron nitride. The position of this feature depends on type of the adsorbate; in the presence of TCNE it is above $E_F^{DFT}$, while in presence of TTF it is below $E_F^{DFT}$. Figures 7b,c, 8b,c, and 9b,c show that this asymmetry leads to a large increase in the thermopower, upon doping.

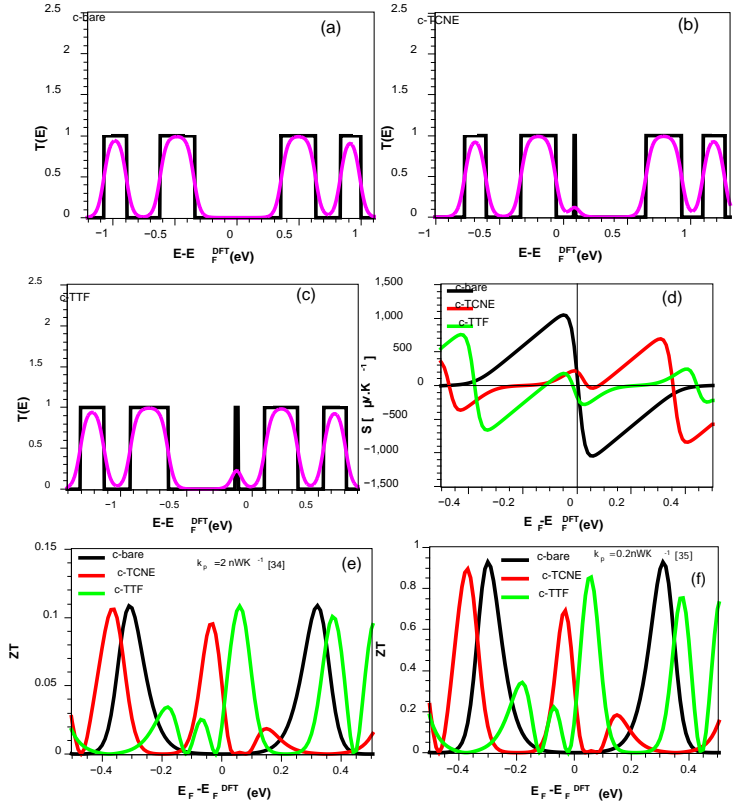

Figure 8: For the graphene-boron nitride structure (c) of figure S8, the black solid line in each panel shows DFT calculations of the transmission coefficient $T(E)$ for electrons of energy $E$ passing from one electrode to the other (a) without doping, (b) doped by TCNE and (c) doped by TTF. The dotted lines show the room temperature electrical conductance. (color online) Figures 8d, 8e and 8f show the room-temperature thermopower $S$, the figure of merit $ZT$, obtained by assuming a phonon thermal conductance of 2nW/K [34] and 0.2nW/K [35]. Since the Fermi energy $E_F^{DFT}$ predicted by density functional theory is not necessarily accurate, Figs (d-f) show results for $S$ and $ZT$ obtained from equ. (4) using different values of $E_F$.

To obtain an estimate of the figure of merit ZT, we need an estimate of the phonon contribution $k_p$ to the thermal conductance in addition to the electronic contribution $k_e$ given by equation (7) and presented in the SI. Thermal conductance can be reduced by both surface roughness and boundary scattering and by nanostructuring ribbons.

Such phonon engineering can take place remotely, in the current –carrying regions which feed electrons into the above heterostructures. Therefore to estimate $k_p$ for the structure of figure 4, we note that a perfect zigzag boron nitride is predicted to have a thermal conductance of 2nW/K[34]. On the other hand, mixed graphene nanoribbons of this width may have a room temperature thermal conductance as low as 0.2 nW/K[35]. As shown in figures 7,8 and 9, after doping with TCNE and TTF the latter leads to a full *ZT* as high as 0.9 at room temeperature, whereas the former leads to values of order 0.1.

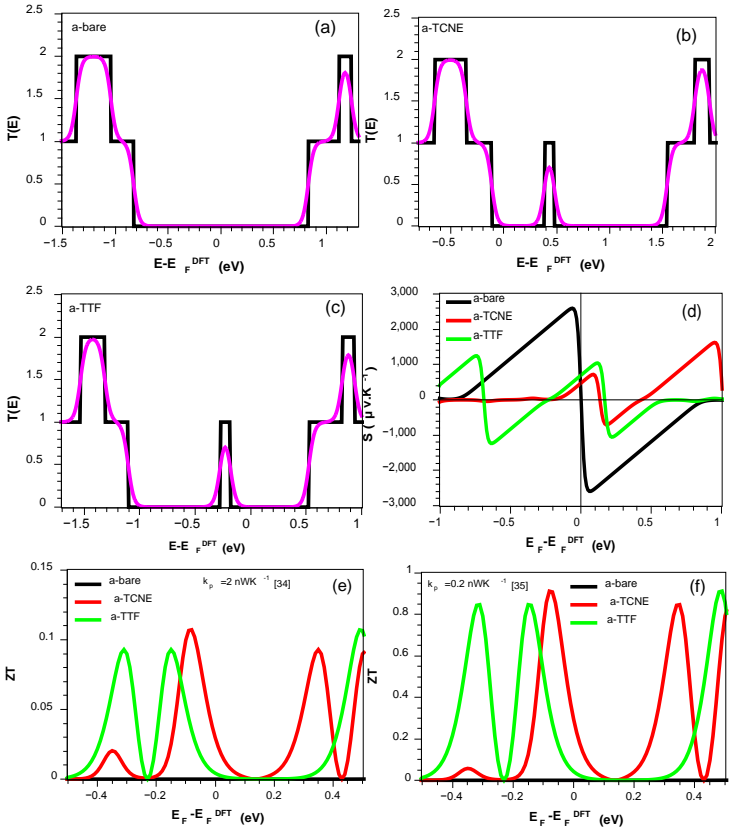

**Figure 9:** For the graphene-boron nitride structure (a) of figures 2-4, the black solid line in each panel shows DFT calculations of the transmission coefficient $T(E_F)$ for electrons of energy $E_F$ passing from one electrode to the other (a) without doping, (b) doped by TCNE and (c) doped by TTF. The dotted lines show the room temperature electrical conductance. (color online) Figures 9d, 9e and 8f show the room-temperature thermopower *S*, the figure of merit *ZT*,

obtained by assuming a phonon thermal conductance of 2nW/K [34] and 0.2nW/K [35].

## 5. Conclusion

We have presented a study of the thermoelectric properties of graphene-boron nitride hetero-structures with various widths of graphene and a fixed width of boron nitride. Our aim was to engineer step-like features in the transmission coefficient *T(E)* located asymmetrically relative to the Fermi energy $E_F$. Although such features arise in the undoped case, they are located symmetrically relative to $E_F$ and do not enhance thermoelectric performance. To induce an asymmetry, we investigated the effect of doping with TCNE and TTF adsorbates. Since TCNE is an electron acceptor, the Fermi energy upon doping must lie between the valence band of the ribbon and the TCNE LUMO. This Fermi-level shift leads to a marked enhancement of both the thermopower and electronic figure of merit. Similarly, since TTF is a donor, the Fermi energy must lie between the conductance band of the ribbon and the TTF HOMO.

The precise value of *ZT* will depend on the phonon thermal conductance of our finite-width wires. Although calculation of phonon transport is beyond the scope of this paper, we can obtain an estimate from recent literature values. For a perfect boron nitride nanowire of the same width[34], the room temperature thermal conductance takes a value of $2nWK^{-1}$, whereas for a structured graphene nanowire wire[35], the room temperature thermal conductance is as low as 0.2 nW $K^{-1}$. The latter is lower than that of a straight graphene wire, due to boundary scattering and similarly we expect the thermal conductance of graphene-boron nitride hetero-structures to be lower than $2nWK^{-1}$ due to interface scattering. The presence of dopants will introduce additional phonon scattering and tend to further reduce the thermal conductance. Taking a phonon thermal conductance of 0.2 nW $K^{-1}$ leads to a *ZT* as high as *ZT=0.9.* Depending on the choice of adsorbate, find that the thermopower could be either positive or negative. When combined with the high value of ZT, this demonstrates that graphene-boron nitride hetero-structures are promising and versatile materials for thermoelectric applications at the nanoscale.

Finally we note that our analysis is based on the Landauer formula, which is valid for ballistic, quasi-ballistic and disordered structures and for any value of the elastic mean free path. The only restriction is that electron transport is elastic. Ie the dimensions of the scattering region should be smaller than the inelastic mean free path. In practice, the inelastic mean free path may be set by the surrounding environment and the temperature, as well as the scattering region of interest. The validity of the Landauer formula was demonstrated in the early 1980's in the context of electron transport through quantum point contacts, which are ballistic region of finite width connected to much wider electrical contacts. For such structures, the transmission coefficient is equal to the number of open scattering channels and the conductance at low temperatures is an integer multiple of the conductance quantum. Atomic wires formed from periodic chains of atoms and clean carbon nanotubes exhibit similar behaviour. The structures we consider are also periodic nanowires and are therefore in the absence of inelastic scattering, the conductance is equal to the number of open channels. The inelastic scattering length is not known theoretically and ultimately should be determined by experiment. It is known that for related nanowires, the inelastic scattering length at room temperature exceeds 3nm[36, 37] and therefore this may set the maximum wire width for the strict applicability of the Landauer formula.

## Acknowledgment

This work is supported by the UK EPSRC, EP/K001507/1, EP/J014753/1, EP/H035818/1, and from the EU ITN MOLESCO 606728. Ministry of Higher Education and Scientific Research, Al Qadisiyah University, IRAQ.